\documentstyle[epsfig,eqsecnum,aps]{revtex}
\newcommand{\be}{\begin{equation}}
\newcommand{\ee}{\end{equation}}
\newcommand{\bea}{\begin{eqnarray}}
\newcommand{\eea}{\end{eqnarray}}
\begin{document}
%\twocolumn
\title{Ground state particle-particle correlations and double
beta decay}
\author{A. A. Raduta$^{a,b)}$, 
P. Sarriguren$^{c)}$, Amand Faessler$^{d)}$ and  E. Moya de Guerra$^{c)}$ }
\address{$^{a)}$Dep. of Theoretical Physics and
Mathematics, Bucharest University, POB MG11, Romania}
\address{$^{b)}$Institute of Physics and Nuclear Engineering, Bucharest, 
POBox MG6, Romania}
\address{$^{c)}$Instituto de Estructura de la Materia, CSIC,
Serrano  119-123, 28006 Madrid, Spain}
\address{$^{d)}$Institut fuer Theoretische Physik der Universitaet Tuebingen, 
Auf der Morgenstelle 14, Germany}
\date{\today}
\maketitle
\begin{abstract}
A self-consistent formalism for the double beta decay of Fermi type is provided.
The particle-particle channel of the two-body interaction is considered first
in the mean field  equations and then in the QRPA. The resulting approach
is called the QRPA with a self-consistent mean field (QRPASMF). The mode 
provided by QRPASMF, does not collapse for any strength of the particle-particle 
interaction. The transition amplitude for double beta decay is almost 
insensitive to the variation of the particle-particle interaction.
Comparing it with the result of the standard $pnQRPA$, it is smaller by a 
factor 6. The prediction for transition amplitude agrees quite well with the 
exact result. The present approach is the only one which produces a strong 
decrease of the amplitude and at the same time does not alter the stability 
of the ground state.  
\end{abstract}
%%%%%%%%%%%%%%%%%%%%%%%%%%%%%%%%%%%%%%%%%%%%%%%%%%%%%%%%%%%%%%%%%%%%%%%
%                          ABSTRACT                                   %
\section{Introduction}
\label{sec: level1}
Many body theories have been very successful in describing the collective 
properties of nuclear systems 
\cite{Kis,Bel,Bay,Chas,Bar,Tok,Ham,Gru,Ike,Marsh,Moy,Rad1}. Many properties 
are described by considering only the two-body interaction between protons 
or between neutrons. However there are several features which cannot be 
explained if one ignores the interactions of protons and neutrons
\cite{Gos,Cam,Jea,Wol,Go1,Lan,Go2,Goo,Tak,Mul,Cheo,Engel,Sat,Tera}. For
example the pairing interaction between protons and neutrons seems to be 
important for determining both the ground state and the high spin states 
properties of nuclei having $Z=N$ \cite{Lan,Goo,Mul,Sat}. The monopole and 
dipole proton-neutron interaction was intensively studied in connection 
with the double beta decay process \cite{Hax,Fae,Sim,Suh}. This subject 
was considered by many theoreticians especially in connection to the 
neutrino-less double beta decay ($0\nu\beta\beta$). Indeed, this phenomenon 
if it exists, might be crucial for answering a fundamental question of 
physics, namely whether the neutrino is a Majorana or a Dirac particle.
In order to make predictions in this field one needs reliable tests for the 
nuclear matrix elements. Unfortunately, up to date, such tests are
missing. However, since essentially the same matrix elements are used for 
the two neutrino double beta decay ($2\nu\beta\beta$) for which plenty of 
data exist, the idea of using those NN interactions, which provide a 
realistic description of the $2\nu\beta\beta$ process has been widely adopted.
The present status of the theories proposed for $2\nu\beta\beta$ is as 
follows: The approach which produces results closest to experimental data 
is the proton-neutron quasiparticle random phase approximation ($pnQRPA$). 
The predictions for the transition amplitude is however too high comparing 
it to the existent data. Therefore one has looked for modifications of the
existent formalism in order to obtain the desired quenching of the 
theoretical result. The first new approach was proposed by Cha \cite{Cha} 
for a related subject. Indeed he noticed that the single beta transition 
rate is very sensitive to varying the strength of the two-body interaction 
in the particle-particle (pp) channel. This idea was adopted by most of 
the groups working in the field of $2\nu\beta\beta$ 
\cite{Vog1,Vog2,Civ1,Klap,Suho}. The results are as follows: The transition 
amplitude is almost insensitive to the variation of the strength, $g_{pp}$, 
with which one multiplies the matrix elements of a realistic force, of the 
pp interaction for a large interval starting from zero. Then the amplitude 
is decreasing quickly reaching zero for   $g_{pp}\approx 1$ which, as a  
matter of fact is very closed to its critical value where the $pnQRPA$ 
breaks down. It should be mentioned that  $g_{pp}=1$ is just what a 
realistic force produces. Of course on the fast down sloping part of the
plot the experimental data is met so one could assume that the problem is 
solved. However, in this region the $pnQRPA$ is not a good approach 
since the results are not stable against adding new correlations. 
A positive feature of the $pnQRPA$ is that the Ikeda sum rule (ISR) is
fully satisfied for any value of $g_{pp}$. Several attempts have been made 
to stabilize the $pnQRPA$ ground state, or to move the value zero of the 
transition amplitude to the unrealistic region. Among the proposed 
approaches are the boson expansion technique (BE) \cite{Rad2,Rad3},
the multiple commutators method (MCM)\cite{Grif,JSu}, the renormalized 
$pnQRPA$ \cite{Toi} and the fully renormalized $pnQRPA$ \cite{Rad5}.
The first three methods succeed to shift the zero of the transition 
amplitude to the unphysical region but the Ikeda sum rule is violated by 
an amount ranging from 15 to 30\%. Some complementary features of the 
boson expansion and re-normalization procedure were studied in 
ref.\cite{Rad6}. Ikeda sum rule is the difference of the total strengths 
for beta minus and beta plus transitions and is influenced by the 
completeness of the states participating in the process as well as by 
the correlations involved in the ground state. Some groups consider Ikeda 
sum rule as a signature for the fact that the formalism takes care of the 
Pauli principle \cite{Sim1}. However, even if the Pauli principle is 
preserved, ISR can be violated if the states involved do not have a good 
particle number.

Very recently we started investigating the consistency of the $pnQRPA$ 
treatment of the pp interaction and the structure of the mean field
\cite{Rad7,Rad8}. Indeed, without exception all previous formalisms
included the particle-particle proton-neutron interaction in the
$pnQRPA$ approach but ignored this interaction when the mean field
was defined. In the previous paper we investigated this problem 
within a consistent approach. We have proved that including the
particle-particle interaction in the single particle mean field
prevents the QRPA to collapse when the interaction strength is
increased, despite the attractive character of the  pp interaction.
Moreover the RPA energy is an increasing function of the interaction 
strength. Here we shall call the procedure mentioned above as the QRPA 
with a self-consistent mean field, QRPASMF.

In the present paper we study the double beta Fermi decay by using the 
wave functions provided by the QRPASMF approach. We are studying the
Fermi transition since this has the nice feature that for a single j 
case, the model is exactly solvable and therefore one can test the 
accuracy of the adopted approximations.  The paper is organized as 
follows: In Section 2 we review briefly the results obtained in the
previous paper. On this occasion we collect the main results needed for 
the present purposes. In Section 3, the QRPA formalism is written for 
the quasiparticle operators which admit  the generalized BCS wave
function as vacuum. As we showed in ref.[46] the generalized BCS wave 
function is the static ground state of the system under consideration. 
The exact treatment is presented in Section IV. The results for the 
four systems involved in the processes considered here, are shown in
the tables. The equations defining the double beta transition amplitude 
as well as the ISR are given in Section V. The numerical results are 
discussed in Section VI while the final conclusions are summarized in 
Section VII. 

\section{Brief review of the classical description of the static
ground state}
\label{sec: level2}
Since the present formalism is a natural continuation of our previous 
work \cite{Rad8}, where the particle- hole (ph) and  particle-particle 
(pp) interactions contribute to the structure of the ground state, it is 
worth summarizing the main achievements from ref.\cite{Rad8}. This will 
help us to define the context of the present subject as well as to fix 
the notations and conventions.

We consider a heterogeneous system of nucleons which move in a spherical 
shell model mean field and interact among themselves in the following 
manner. Alike nucleons interact through monopole pairing forces while 
protons and neutrons interact by a monopole particle-hole  and a monopole
particle-particle  two-body term.
Such a system is described by the following many-body Hamiltonian:
\begin{eqnarray}
H&=&\sum_{\tau,j,m}(\epsilon_{\tau}-\lambda_{\tau})c^{\dag}_{\tau jm}c_
{\tau jm}  \nonumber \\
&-&
\frac{G_{p}}{4}\sum_{j,m;j^{\prime},m^{\prime}} c^{\dag}_{pjm}
c^{\dag}_{\widetilde{pjm}}
c_{\widetilde{pj^{\prime}m^{\prime}}}c_{pj^{\prime}m^{\prime}}
-\frac{G_{n}}{4}
\sum_{j,m;j^{\prime},m^{\prime}} c^{\dag}_{njm}c^{\dag}_{\widetilde{njm}}
c_{\widetilde{nj^{\prime}m^{\prime}}}c_{nj^{\prime}m^{\prime}}
\nonumber\\
&+&2{\chi}(c^{\dag}_{pj}c_{nj})_{0}(c^{\dag}_{nj^{\prime}}c_{pj^{\prime}})_{0}
 -2\chi_1(c^{\dag}_{pj}c^{\dag}_{nj})_{0}
(c_{nj^{\prime}}c_{pj^{\prime}})_{0}.
\end{eqnarray}
$c^{\dag}_{\tau jm}(c_{\tau jm})$ denotes the creation (annihilation) operator
of one particle of $\tau$($=p,n$) type in the spherical shell model state
$|\tau;nljm\rangle\equiv|\tau jm\rangle$.  The time reversed state
corresponding to  $|\tau;nljm\rangle$ is denoted by
$|\widetilde{\tau;jm}\rangle=(-)^{j-m}|\tau;j-m\rangle$.
Several authors used this Hamiltonian to describe single and double beta Fermi 
transitions within a $pnQRPA$ formalism\cite{Sim1,Rad7,Rad8,Sam}.

The static ground state of this Hamiltonian is described by the stationary 
solutions of the time dependent equations of motion provided by the 
variational principle\cite{Moy,Rad1,Vil}:
\be
\delta \int_{0}^{t}\langle\Psi|H-i\frac{\partial}{\partial t^{\prime}}
|\Psi\rangle dt^{\prime} =0, 
\ee
with the trial function  $|\Psi\rangle$  having  the following form:
\begin{equation}
|\Psi\rangle=\Psi(z_{p}, z^{*}_{p}; z_{n}, z^{*}_{n}; z_{pn}, z^{*}_{pn})=
e^{T_{pn}}e^{T_{p}}e^{T_{n}}|0\rangle.
\end{equation}
The transformations specified by the operators T are given by:
\begin{eqnarray}
T_{pn}&=&\sum_{jm}(z_{pnj}c^{\dag}_{pjm}c^{\dag}_{\widetilde{njm}}-
z^{*}_{pnj}c_{\widetilde{njm}}c_{pjm}), \\
T_{p}&=&\sum_{jm}(z_{pj}c^{\dag}_{pjm}c^{\dag}_{\widetilde{pjm}}-
z^{*}_{pj}c_{\widetilde{pjm}}c_{pjm}), \\
T_{n}&=&\sum_{jm}(z_{nj}c^{\dag}_{njm}c^{\dag}_{\widetilde{njm}}-
z^{*}_{nj}c_{\widetilde{njm}}c_{njm}).
\end{eqnarray}
$|0\rangle$ stands for the particle vacuum state. These transformations 
depend on the parameters $z$ which are complex functions of time. The 
corresponding complex conjugate functions are denoted by $z^*$. The 
parameters ($z,z^*$) play the role of classical coordinates and conjugate 
momenta, respectively. The classical coordinates are related to the usual 
coefficients $U,V$ entering the Bogoliubov-Valatin (BV) transformations 
by the equations:
\bea
z_{pj}& = &\rho_{pj}e^{i\varphi_{pj}},\; U_{pj}=\cos2\rho_{pj},\;
V_{pj}=e^{-i\varphi_{pj}}\sin2\rho_{pj},\nonumber\\ 
 z_{nj}& = &\rho_{nj}e^{i\varphi_{nj}},\;
U_{nj}=\cos2\rho_{nj},\;
V_{nj}=e^{-i\varphi_{nj}}\sin2\rho_{nj},\nonumber\\ 
z_{pnj}& = &\rho_{pnj}e^{i\varphi_{pnj}},\;
U_{j}=\cos\rho_{pnj}, \;   
V_{j}=e^{-i\varphi_{pnj}}\sin\rho_{pnj}.
\eea
Denoting by ${\hat U}$ the transformation (2.3) which transforms the bare 
vacuum into the trial function $|\Psi\rangle$, and by $\alpha^\dagger$ the 
images of the particle creation operators through this transformation one 
easily proves that:
\bea
\left(\matrix{\alpha^{\dag}_{1jm}\cr \alpha^{\dag}_{2jm}
\cr \alpha_{\widetilde{1jm}}\cr \alpha_{\widetilde{2jm}}}\right)=
\left(\matrix{U_{pj}U_j & -V_{pj}V_j^* & -V_{pj}U_j & -U_{pj}V_j \cr
              -V_{nj}V^*_j & U_{nj}U_j & -U_{nj}V_j &-V_{nj}U_j \cr
              V^*_{pj}U_j & U_{pj}V^*_j & U_{pj}U_j &-V^*_{pj}V_j \cr
              U_{nj}V^*_j & V^*_{nj}U_j & -V^*_{nj}V_j &U_{nj}U_j}\right)
\left(\matrix{c^{\dag}_{pjm}\cr c^{\dag}_{njm}
\cr c_{\widetilde{pjm}}\cr c_{\widetilde{njm}}}\right).
\eea
The static ground state of the model Hamiltonian corresponds to
a minimum value for the classical energy function:
\bea
{\cal H}&=&\sum_{j}(\epsilon_{pj}-\lambda_{p})(2j+1)V^{2}_{{\rm eff},pj}+
\sum_{j}(\epsilon_{nj}-\lambda_{n})(2j+1)V^{2}_{{\rm eff},nj}
\nonumber\\
&&-\frac{|\Delta_{p}|^{2}}{G_{p}}-\frac{|\Delta_{n}|^{2}}{G_{n}}+	   
\frac{2|\beta_{-}|^{2}}{\chi}-\frac{2|\Delta_{pn}|^{2}}{\chi_{1}},
\eea
where the following notations have been used:

\begin{eqnarray}
V^{2}_{{\rm eff},pj}&=&(U_{j}^{2}|V_{pj}|^2+|V_{j}|^{2}U_{nj}^2),\nonumber\\
V^{2}_{{\rm eff},nj}&=&(U_{j}^{2}|V_{nj}|^2+|V_{j}|^{2}U_{pj}^2),\nonumber\\
\Delta_{p}&\equiv&\frac{G_{p}}{2}\langle\Psi|\sum_{j,m} c^{\dag}_{pjm}
c^{\dag}_{\widetilde{pjm}}|\Psi\rangle=\frac{G_{p}}{2}\sum_{j}(U^{2}_{j}U_{pj}
V_{pj}-V_{j}^{2}U_{nj}{V_{nj}}^{*}),\nonumber\\
\Delta_{n}&\equiv&\frac{G_{n}}{2}\langle\Psi|\sum_{j,m}c^{\dag}_{njm}
c^{\dag}_{\widetilde{njm}}|\Psi\rangle=\frac{G_{n}}{2}\sum_{j}(U^{2}_{j}U_{nj}
V_{nj}-V_{j}^{2}U_{pj}{V_{pj}}^{*}),\nonumber\\
\beta_{-}&\equiv&\chi\langle\Psi|\sum_{j}(c^{\dag}_{pj}c_{nj})_{0}|\Psi
\rangle= \chi\sum_{j}\hat j(U_{j}V_{j}^{*}U_{pj}V_{pj}+U_{j}V_{j}U_{nj}
{V_{nj}}^{*}),\nonumber\\
\beta_{+}&\equiv&(\beta_{-})^{*}=\chi\langle\Psi|
\sum_{j}(c^{\dag}_{nj}c_{pj})_{0}|\Psi\rangle,\nonumber\\
\Delta_{pn}&\equiv&\chi_{1}\langle\Psi|\sum_{j}(c^{\dag}_{pj}c^{\dag}_{nj})_{0}
|\Psi\rangle=\chi_{1}\sum_{j}\hat jU_{j}V_{j}(U_{pj}^{2}-|V_{nj}|^{2}),
\;{\hat j}=\sqrt{2j+1}.
\end{eqnarray}
The Fermi level energies $\lambda_p$ and $\lambda_n$ are determined so
that the average number of protons and neutrons are equal to $Z$ and $N$, 
respectively:
\bea
Z=\sum_{j}(2j+1)V_{{\rm eff},pj}^2,\nonumber \\
N=\sum_{j}(2j+1)V_{{\rm eff},nj}^2.
\eea
As in ref.[46] we restrict our considerations to the single j case. In the 
paper quoted above we determined the minimum points of the classical energy 
${\cal H}$, by solving the generalized BCS equations consisting of four
equations for the gap parameters $\Delta_p, \Delta_n,\Delta_{pn}$ and 
$\beta_-$ and two constraint equations given by (2.11). The solutions 
determine the variables $V_p,V_n, V$ which by means of (2.7) and (2.3) 
provide the static ground state. The next step achieved in ref. [46] was the 
description of the small oscillations of the system around the static ground 
state. The procedure is equivalent to an RPA formalism. However the 
semi-classical approach used there requires a certain caution when is 
applied to transition amplitudes. The reason is that while the RPA 
approximation is a quadratic expansion of the energy function around its 
minimum point, the classical counterpart of transition amplitudes may 
achieve their minima in some points of the phase space, different from the
energy minimum. In this case the transition amplitudes are to be either 
expanded around their own minima or around the energy minimum point. In the 
first case it is difficult to perform a consistent quantization of both 
energies and transition amplitudes while in the second case truncating 
the expansion at its second order might be not sufficient. In order to 
avoid these complications, here we adopt the quasiparticle random phase 
approximation (QRPA) with respect to the quasiparticle operators having 
the static ground state determined so far, as a vacuum state.
The new method will be described in detail in the next Section.  

\section{QRPASMF formalism for the single j case}
\label{sec: level 3}
Inserting the coefficients U and V determined by the pairing equations 
into Eq. (2.8) one obtains a generalized Bogoliubov-Valatin transformation 
for the quasiparticle operators $\alpha^\dagger_1,\alpha^\dagger_2$. 
Reversing the BV transformation (2.8) and denoting by T the
resulting matrix transformation, one obtains: 

\bea
\left(\matrix{c^{\dag}_{pjm}\cr c^{\dag}_{njm}
\cr c_{\widetilde{pjm}}\cr c_{\widetilde{njm}}}\right) =
\left(\matrix{T_{p1}&T_{p2} &T_{\widetilde{p1}} &  T_{\widetilde{p2}}   \cr
                   T_{n1}&T_{n2} &T_{\widetilde{n1}} & T_{\widetilde{n2}} \cr 
-T^*_{\widetilde{p1}} &-T^*_{\widetilde{p2}}&T^*_{p1}&T^*_{p2}\cr 
-T^*_{\widetilde{n1}} & -T^*_{\widetilde{n2}}&T^*_{n1}&T^*_{n2} }\right) 
\left(\matrix{\alpha^{\dag}_{1jm}\cr \alpha^{\dag}_{2jm}
\cr \alpha_{\widetilde{1jm}}\cr \alpha_{\widetilde{2jm}}}\right).
\eea

Making use of this transformation we can easily express the
single j restriction of the model Hamiltonian in the
quasiparticle representation
\be
H_{qp}=E_{11}\alpha^\dagger_1\alpha_1+E_{22}\alpha^\dagger_2\alpha_2+
E_{12}\alpha^\dagger_1\alpha_2+E_{21}\alpha^\dagger_2\alpha_1
+4\chi V^\dagger_{ph}V_{ph}-4\chi_1 V^\dagger_{pp}V_{pp}.
\ee
The quasiparticle energies are given analytically in the Appendix A. 
The two-body terms are written in a factorized form. Ignoring the scattering 
terms, since they do not contribute to the QRPA equations, the factors
involved are linear combinations of monopole two quasiparticle operators:
\bea
V^\dagger_{ph}=&&T_{p1}T^*_{\widetilde{n1}}A^\dagger_{11}+
\frac{1}{\sqrt{2}}(T_{p1}T^*_{\widetilde{n2}}+T_{p2}T^*_{\widetilde{n1}})
A^\dagger_{12}+T_{p2}T^*_{\widetilde{n2}}A^\dagger_{22} \nonumber \\
&&+T_{\widetilde{p1}}T^*_{n1}A_{11}+\frac{1}{\sqrt{2}}
(T_{\widetilde{p2}}T^*_{n1}+T_{\widetilde{p1}}T^*_{n2})A_{12}+
T_{\widetilde{p2}}T^*_{n2}A_{22}, \nonumber \\
V^\dagger_{pp}=&&T_{p1}T_{n1}A^\dagger_{11}+
\frac{1}{\sqrt{2}}(T_{p1}T_{n2}+T_{p2}T_{n1})
A^\dagger_{12}+T_{p2}T_{n2}A^\dagger_{22} \nonumber \\
&&-T_{\widetilde{p1}}T_{\widetilde{n1}}A_{11}-\frac{1}{\sqrt{2}}
(T_{\widetilde{p1}}T_{\widetilde{n2}}+T_{\widetilde{p2}}T_{\widetilde{n1}})
A_{12}-T_{\widetilde{p2}}T_{\widetilde{n2}}A_{22}.
\eea
The following notations for two quasiparticle operators are used: 
\bea
A^\dagger_{kk}&=&\frac{1}{\sqrt{4\Omega}}\sum_{m}\alpha^{\dagger}_{kjm}
\alpha^{\dagger}_{\widetilde{kjm}},~k=1,2 \nonumber\\
A^\dagger_{12}&=&\frac{1}{\sqrt{2\Omega}}\sum_{m}\alpha^{\dagger}_{1jm}
\alpha^{\dagger}_{\widetilde{2jm}}.
\eea
For a compact writing of the forthcoming equations, it is convenient
to introduce the following notations:
\be
\Gamma^\dagger_1=A^{\dagger}_{11},~\Gamma^\dagger_2=A^{\dagger}_{12},
~\Gamma^\dagger_3=A^{\dagger}_{22}.
\ee
Assuming the quasi-boson approximation for the commutators of the
operators $\Gamma^\dagger_k$ and their hermitian conjugate, one
obtains the following set of linear equations of motion:
\bea
\left[H,\Gamma^{\dagger}_k\right ]&=&
\sum_{l}\left(A_{kl}\Gamma^{\dagger}_l+ B_{kl}\Gamma_l\right),
\nonumber\\
\left[H,\Gamma_k\right ]&=&
\sum_{l}\left(-B_{kl}\Gamma^{\dagger}_l- A_{kl}\Gamma_l\right).
\eea
where the matrices $A$ and $B$ are defined in the Appendix B. 
The QRPA approximation determines a boson operator $C^{\dagger}$
as a linear combination of the $\Gamma^{\dagger}$ and $\Gamma$
which describes a harmonic motion of the many-body system:
\bea
C^{\dagger}&=&\sum_{k}\left(X_k\Gamma^{\dagger}_k-Y_k\Gamma_k\right),\\
\left[H,C^{\dagger}\right]&=&\omega C^{\dagger},\\
\left[C,C^{\dagger}\right]&=&1.
\eea 
Equations (3.8) supply us with the QRPA equations for the
phonon amplitudes X and Y:
\bea
\left(\matrix{A&B \cr -B&-A}\right)
\left(\matrix{X\cr Y}\right)=\omega \left(\matrix{X\cr Y}\right),
\eea
while the boson condition (3.9) provides the normalization equation:
\be
\sum_{k}\left(|X_k|^2-|Y_k|^2\right)=1.
\ee 
\section{Exact treatment}
\label{sec: level4}
Restricting the model Hamiltonian to a single j shell, the resulting 
many-body Hamiltonian can be exactly treated. Indeed, its eigenvalues 
may be found through a diagonalization procedure in a many-body basis. 
Since in our application we consider a system of 12 neutrons and 4 
protons distributed in  shells of equal angular momenta ($\frac{19}{2}$), 
and which may decay through the channels of $\beta^+,\beta^-$ and 
$2\nu\beta\beta$, the model Hamiltonian is to be considered for the 
following 4 systems with $(N,Z)=(13,3),(12,4),(11,5),(10,6)$, respectively.
In what follows we shall refer to the above mentioned systems as
to the grand mother, mother, intermediate and daughter nuclei,
respectively. To specify the paternity of the many-body basis,
they will be accompanied by the indices (gm), (m), (i), (d), respectively.
Here the following basis, formed out of non-orthogonal states, will be used:
\be
|n_1,n_2,n_3\rangle={\cal N}_{n_1n_2n_3}
(A^{\dagger}_{pp})^{n_1}(A^{\dagger}_{nn})^{n_2}
(A^{\dagger}_{pn})^{n_3}|0\rangle
\ee
where $A^{\dagger}_{pp},A^{\dagger}_{nn},A^{\dagger}_{pn}$ 
are monopole two-particle excitation operators:
\bea
A^{\dagger}_{pp}=\sum_{m}c^{\dagger}_{pjm}
c^{\dagger}_{\widetilde{pjm}}, \\ \nonumber
A^{\dagger}_{nn}=\sum_{m}c^{\dagger}_{njm}c^{\dagger}_{\widetilde{njm}},
\\ \nonumber
A^{\dagger}_{pn}=\sum_{m}c^{\dagger}_{pjm}c^{\dagger}_{\widetilde{njm}}.
\eea
The powers $n_1,n_2,n_3$ appearing in the defining Eq. (4.1)
are different for the four nuclei mentioned above. Indeed they are
subject to the restrictions:
\be
2n_1+n_3=Z,~2n_2+n_3=N.
\ee
The solutions of the above equations for integer numbers define the
basis states for the nuclei involved in the transitions
considered in the present paper. These are given in Table 1.

The states are normalized to unity. The normalization factors,
denoted by ${\cal N}_{n_1n_2n_3}$, are listed in Table 2 while
the overlaps of states are collected in Table 3.

In order to find the eigenvalues of the model Hamiltonian H in the
non-orthogonal basis we have to calculate its matrix elements.
This goal is achieved in Appendix C. Let us denote by ${\bar H}$
the resulting matrix associated to H. Also we denote by $O$ the
overlap matrix. As we showed in ref. [46], the eigenvalue equation: 
\be
H|\Phi\rangle_l=|\Phi\rangle_l,~l=gm,m,i,d,
\ee
with
\be
|\Phi\rangle_l=\sum_{k}C_k|k\rangle_l
\ee
can be reduced to an ordinary eigenvalue problem for a symmetric matrix:
\be
{\widetilde{H}}X=EX,
\ee
where
\be
\widetilde{H}=U^{-1}{\bar H}(U^{-1})^T,~X=U^TC,
\ee
while $U$ is determined by the factorization:
\be
O=UU^T.
\ee
Therefore the eigenvalues $E$ are determined by a standard diagonalization 
procedure for the symmetric matrix $\widetilde{H}$ and the eigenfunction 
$|\Phi\rangle$ is determined by the column vector
\be
C=(U^T)^{-1}X.
\ee
These results will be used in the next Section to calculate the transition 
amplitude for the double beta Fermi transition. The dependence of the 
energies obtained for the mother, intermediate and daughter nuclei, on 
the strength parameter of the pp interaction is shown in Fig. 1.
\section{Double beta transition.}
\label{sec: level5}
The double beta decay with two neutrinos in the final state is considered 
to consist of two consecutive single $\beta^-$ decays. The intermediate 
state, reached by  the first $\beta^-$ transition, consists of an 
odd-odd nucleus in a $pn$ excited state, one electron and one 
anti-neutrino. If  in the intermediate state, the total lepton energy
is approximated by the sum of the electron rest mass and half of
the Q-value of the double beta decay process, the inverse of the
process half-life can be factorized as follows:
\be
(T^{2\nu}_{1/2})^{-1}=F|M_F|^2,
\ee
where F is a lepton phase integral while the second factor
is determined by the states characterizing the nuclei involved
in the process and is given by  the expression:
\be
M_F=\sum_{k, k^{\prime}}\frac{{_m}\langle {0}^+|{\beta}^+|{0^+}_k{\rangle}_m 
{_m}\langle 0^+_k| 0^+_{k'}{\rangle }_d
{_d}\langle 0^+_{k'}|{\beta }^+|0^+{\rangle}_d}{E_{k}+\Delta E}
\ee
where the transition operator is
\be
{\beta}^+=\sum_{m}c^{\dagger}_{njm}c_{pjm}
\ee
The initial and final states are the ground states of the mother
$(|0^{\dagger}\rangle _m)$ and daughter $(|0^{\dagger}\rangle_d)$
nuclei. In the QRPASMF formalism these are described by the phonon
operator vacuum state while in the exact treatment these are the
lowest states obtained through diagonalization, respectively.
The intermediate states are one phonon states in the QRPASMF approach 
while in the exact treatment they are eigenstates of the model Hamiltonian
describing the odd-odd system. The corresponding eigenvalues,
normalized to the ground state energy of the mother nucleus, are
denoted by $E_k$. Since the excited states corresponding to the two 
vacua are not orthogonal to each other the overlap matrix elements should 
be inserted between the matrix elements describing the two legs of the 
double beta transition. For the exact treatment the overlap of the 
eigenstates is obtained by multiplying the vectors of the weights produced 
by diagonalization with the overlap matrix. In the QRPASMF treatment,
the overlap matrix elements are:
\be
{_m}\langle {0^+}_k| {0^+}_{k'}{\rangle}_d= X^m_{1k}X^d_{1k'}+X^m_{
2k}X^d_{2k'}+X^m_{3k}X^d_{3k'}-Y^m_{1k}Y^d_{1k'}
-Y^m_{2k}Y^d_{2k'}-Y^m_{3k}Y^d_{3k'}.
\ee
The matrix elements describing single $\beta^+$ and $\beta^-$
transitions of the mother system satisfy the  Ikeda sum rule:
\be
\beta^{(-)}-\beta^{(+)}=N-Z,
\ee
where $\beta^{(-)}$ and $\beta^{(+)}$ denotes the total strength
for $\beta^-$ and $\beta^+$ transitions respectively, while N, Z
specifies the number of neutrons and protons.
\section{Numerical results}
\label{sec: level6}
We applied the formalism described in the previous sections to a
system of 12 neutrons and 4 protons moving in a shell with
$j=\frac{19}{2}$. We study the decay of this system to the daughter 
nucleus with $(N,Z)=(10,6)$. This transition is achieved in two steps, 
with the intermediate odd-odd system having $(N,Z)=(11,5)$. The static 
solution for the equations of motion were found using the following set 
of parameters for the model Hamiltonian restricted to the case of a 
single shell:
\be
\epsilon_p=\epsilon_n=3MeV,~G_p=0.25MeV,~ G_n=0.12 MeV,~\chi=0.20 MeV.
\ee
The strength of the pp interaction, $\chi_1$, is varied freely
in the interval [0.0,1.5]MeV. For each value of $\chi_1$, we solved the 
generalized BCS equations for both mother and daughter nuclei. As a 
result one obtains the gaps and chemical potentials and subsequently the
U and V coefficients. In the next step, the QRPA equation is solved. 
One should remark that, as shown in ref. [46], there are two constants of
motion which results in having two spurious solutions. Therefore there 
exists only one physical solution which corresponds to the matrix 
elements $A_{11}$ and $B_{11}$, i.e. the degree of freedom accounted 
by the amplitudes $X_1, Y_1$. The RPA energies and phonon amplitudes 
are used then to calculate the amplitude for the double beta decay, 
given by eq.(5.2). The results for $M_F$ are given, as function of
$\chi_1$, in Fig. 2. The present QRPA approach is different from the 
standard $pnQRPA$, where the phonon operator is a neutron-hole
proton-particle excitation. Here, due to the proton-neutron
mixing in the quasiparticle operator $\alpha_{1jm}$, the phonon operator
comprises also pn scattering terms $(c^{\dag}_pc^{\dag}_n)$ as well as 
charge conserving excitation operators $(c^{\dag}_pc_{p^{\prime}})$.  
Therefore one expects to obtain a description which is substantially 
different from the one provided by  the standard $pnQRPA$. 
The double beta transition amplitude has been alternatively calculated  by
using the exact eigenstates of the model Hamiltonian H. Thus, H (2.1)
was diagonalized for the three systems involved in the process,
i.e. the mother, intermediate and daughter nuclei. The corresponding 
energies are plotted in Fig. 1 as function of $\chi_1$, for vanishing 
chemical potentials. Several remarks are worth to be mentioned. Note 
that due to the specific model space, $j=\frac{19}{2}$, and the
proton and neutron numbers the mother and intermediate nuclei, have three 
different $0^+$ states while the daughter nucleus exhibits four independent
states. The ground state of the mother nucleus gets higher in energy than 
the ground state of the intermediate nucleus. This happens for $\chi_1$ 
equal to about 0.9 MeV. It is not clear if the zero of the double
beta transition amplitude in the standard $pnQRPA$ approach, at
this value of $\chi_1$, is caused to some extent by this level crossing.
The single beta transition to the odd-odd system is therefore
virtual up to the intersection point and real from there on.
As shown in Fig. 2, for the value of $\chi_1$ where the ground
states of the two neighboring nuclei, mother and intermediate,
are equal to each other, the transition amplitude has a minimum
value. Increasing further the strength $\chi_1$ the amplitude
is slowly increasing and finally reaches a plateau at about $\chi_1=1.6$.
The agreement between the QRPASMF result and the exact one is quite
good. The amplitude predicted by the QRPASMF formalism is varying
very slowly with $\chi_1$ which, in fact, reflects  the stability
of the mean field with respect to the particle-particle
interaction. Note that within QRPASMF the Fermi sea energies $\lambda_p$ and
$\lambda_n$ depend on $\chi_1$, due to the adopted variational procedure.
The exact result shown in Fig. 2, corresponds to those $\lambda_p, \lambda_n$
provided by $QRPASMF$ for $\chi_1=1$.
 We checked that the Fermi energies are only slightly depending
on $\chi_1$ and such a dependence does not  essentially affect
the exact results.

The previous theoretical investigations mostly focused on the
decreasing part of the double beta decay amplitude within  the standard 
$pnQRPA$ description, since the stable result is usually larger than the 
experimental one. Unfortunately the predictions in this region are not 
very stable and moreover show  a zero for a certain $\chi_1$.
In this context it is remarkable that the present formalism
produces the desired moderate suppression of the transition
amplitude and moreover the result is stable against adding more
correlations. 

The T=1 proton-neutron interaction was also considered in refs. [23] and [49] 
to define a generalized pairing field. The formalism was employed for the 
description of the Gamow-Teller $2\nu\beta\beta$ process. However the 
particle-particle two body interaction, whose strength causes the ground 
state instability is of dipole-dipole type and not included in the mean 
field. Accounting also for this dipole-dipole proton-neutron interaction 
requires a simultaneous treatment of the T=1 and T=0 proton-neutron pairing
interactions. Note that such a problem does not appear for the
Fermi double beta transition. Indeed here there is no particle-
particle interaction term which is left out when the mean field
is defined. Extension to the more complex situation of the
Gamow-Teller double beta decay is under work and the results
will be published in a forthcoming paper. 

Let us compare the present results with those given by the standard $pnQRPA$: 
We recall that within the standard $pnQRPA$, used widely in the
literature, first one treats the proton-proton and neutron-neutron pairing 
interactions by the usual BCS procedure (apart from the work of the 
Tuebingen group, which, as we mentioned before, included also
proton-neutron pairing \cite{Cheo,Cheo1}.
In the quasiparticle representation, after ignoring the
scattering terms the Hamiltonian reads:
\be
H=E_p{\hat N}_p+E_n{\hat
N}_n+\lambda_1A^+A+\lambda_2({A^+}^2+A^2), 
\ee
where ${\hat N}_p$ and ${\hat N}_n$ denote the proton and
neutron quasiparticle number operators, respectively. The quasiparticle
energies $E_p$ and $E_n$ have simple expressions:
\be
E_{\tau}=\frac{G_{\tau}\Omega}{2},~\tau=p,n.
\ee
The proton neutron two quasiparticle monopole operator is denoted by:
\be
A^{\dagger}=\frac{1}{\sqrt{2\Omega}}
\sum_{m}a^{\dagger}_{pjm}a^{\dagger}_{\widetilde{njm}}. 
\ee
The coefficients $\lambda_1$ and $\lambda_2$ have simple expressions in 
terms of particle-hole and particle-particle interaction strengths.
\bea
\lambda_1&=&2\chi(U_p^2V_n^2+U_n^2V_p^2)-2\chi_1(U_p^2U_n^2+V_p^2V_n^2),
\\ \nonumber
\lambda_2&=&2(\chi+\chi_1)U_pU_nV_pV_n.
\eea
The $pnQRPA$ energy can be expressed in terms of the
strengths parameters $\lambda_1,\lambda_2$, as
\be
\omega=\sqrt{(E_p+E_n+\lambda_1)^2-4\lambda_2^2}.
\ee
The transition amplitude (5.2) corresponding to this mode  is represented 
in Fig. 3 as a function of $\chi_1$ and will be hereafter referred as the 
standard $pnQRPA$ transition amplitude. Comparing the curves for $M_F$ 
given in Fig. 2 and Fig. 3, one notices that the present formalism predicts 
an amplitude which is smaller by a factor 6 than the amplitude at the stable 
plateau at small $\chi_1$, given by the standard $pnQRPA$. Since in the 
three cases considered here, the exact treatment, the new QRPA and the 
standard $pnQRPA$, we consider the same energy shift ($\Delta E=2.5MeV$) 
in the denominator of eq. (5.2) we have to pay attention to possible 
different zero point energies in the mother and daughter nuclei. 
What is the reason for such a dramatic change in  the magnitude
of the transition amplitude? To see that, we plotted in Fig. 4 the
ratios of the matrix elements describing the two legs of the
double beta transition (see eq. (5.2)), obtained by the standard
$pnQRPA$ and the QRPASMF approach. From there one sees that while the 
$\beta^-$ decay matrix of the first leg elements predicted by the two 
approaches are about the same, the matrix elements describing the second 
decay differ from each other by a factor of about six, in the beginning of 
interval. Although the matrix element describing the second decay in the
present formalism is small it is only slightly changed by
increasing the parameter $\chi_1$.  
Since the static ground state achieves the minimum for the
classical energy one expects that the present RPA approach is a
very good many-body approach. This is confirmed by the results
shown in Fig. 5 where indeed the phonon backward amplitude $Y$ does
not exceed 10$\%$ of the phonon forward amplitude in a large
interval of $\chi_1$. By contradistinction within the standard
$pnQRPA$ approach the Y amplitude for the daughter nucleus
becomes equal to the X amplitude (see Fig. 6), but of opposite sign, 
at $\chi_1$ about 1. For the mother nucleus the equality for the magnitudes 
of the two amplitudes is reached at $\chi_1=1.2$.
At the points mentioned above the $pnQRPA$ phonons collapse,
respectively. The relationship of the two amplitudes suggests
that the phonon operators become hermitian for this strength of
the pp interaction, which prevents a well defined boson
excitation. Note that this bad behavior does not show up in our
formalism. The reason is that the pp interaction is
considered in the definition of the mean field.
The Ikeda sum rule ( ISR) is shown in Fig. 7. We notice that the
predicted ISR differs from the $N-Z$ value by 25$\%$ for 
$\chi_1=1$. Of course the exact result agrees perfectly with the
ISR. A possible reason for the discrepancy quoted above 
could be that in the present approach, the mother nucleus can decay by 
$\beta^-$ and by $\beta^+$ only to one state in each case. By contrast in the
exact treatment the ground state of the mother nucleus is linked
by non-vanishing matrix elements of $\beta^-$ to three states in the
intermediate odd-odd nucleus and by $\beta^+$ to two states in the so called
grandmother nucleus. 

Before closing this Section we would like to comment the comparison of the 
three methods, exact, QRPASMF, pnQRPA, from a different view angle.
Note that the QRPASMF result is depending very little on $\chi_1$
and, in a way, is averaging the behaviour of the exact result
around the value of $\chi_1$ where the relative position of the
ground states of mother and intermediate nuclei is changed.
The monotonic behaviour of $M_F$ is consistent with the QRPASMF
energy represented as function of $\chi_1$. The common feature,
monotony, is caused by the fact that for each value of $\chi_1$ one
determines the corresponding static ground state. Despite the
fact the particle-particle proton-neutron interaction is
attractive the QRPASMF mode does not collapse since, at the same
time, the interaction increases the pairing  energy gap and in
this way the mode energy becomes an increasing
function.
In order to get a deeper understanding of the mechanism causing
the fact that the $M_F$ value predicted by QRPASMF does not
exhibit the allure of the similar function produced by the
exact calculation, some additional comments are necessary.
Indeed, from Fig. 1 it results that for a certain value of $\chi_1$
the ground states for the mother and intermediate nuclei have
the same energy. However we cannot speak about a ground state
degeneracy which might be related to a phase transition for the
nuclear system since we deal with two distinct systems
characterized by different number of protons and neutrons in the
framework of the exact treatment. Note that this level crossing
point produces a minimum value for $M_F$.
On the other hand in the standard pnQRPA formalism, where one
treats first the proton-proton and neutron-neutron pairing
interactions, the small oscillations are treated with respect to
a static ground state which is independent of $\chi_1$. Given
the attractive character of the particle-particle interaction
the mode energy is continuously decreasing and fatally reaches
the value zero. At this point the ground state of the daughter
nucleus becomes degenerate. This reflects the fact that the
static ground state has two dominant components of the same magnitude.
In contradistinction to what happens in  the exact calculations,
the components whose amplitudes become of equal magnitudes are
both associated to even-even systems. Moreover, in the non-exact 
treatment the two degenerate states characterise the same superfluid 
system. In this respect the monotonic decreasing behaviours of the 
$M_F$ functions corresponding to the exact and the pnQRPA descriptions 
are caused by different physical circumstances. In the present formalism,
when one passes from the mother to the intermediate nucleus by
subtracting a neutron and adding a proton one obtains a state
of two quasiparticles. Due to the presence of an energy gap
caused by the proton-neutron pairing interaction the two
systems, the mother and intermediate odd-odd nuclei, will never
have equal energies. Moreover the above mentioned gap is an
increasing function of $\chi_1$ which implies an overall repulsive
character of the $\chi_1$-interaction with respect to this type
of excitation. Aiming at removing possible confusions we stress the fact
that the "exact solution" in the present paper has different
meaning than in all previous publications (see for example Refs [44,48]). 
Indeed  our exact solutions are eigenstates of the many body Hamiltonian 
(2.1) while all previous publications deal with the exact eigenstates of the
quasiparticle Hamiltonian (6.2) which is a severely truncated image of
the initial Hamiltonian (2.1) through the Bogoliubov Valatin transformation.

Concerning the extension of the formalism to a realistic
interaction and a large model space for single particle states
the amount of work involved can be evaluated as follows. For the Fermi
transition  the extension is straightforward and no principle difficulties
appear. As for the Gamow-Teller transition there are
two ways to cope with this problem. a) To treat simultaneously the
T=1 and T=0 pairing and project out the good particle number,
angular momentum and isospin from the resulting generalized BCS
wave function. The projected states are to be used in a time
dependent variational equation to define the QRPA states.
b) A deformed single particle mean field can be defined, as usual, by
the spherical shell model term and the particle-particle proton-neutron
two body interaction. At the second step the pairing interaction
for the deformed single particle states is treated. Finally the 
usual QRPA procedure accounts for quasiparticle correlations.
We would like to mention the fact that there is work in
progress on these lines.
The two extensions mentioned above have in common with the
formalism of the present paper, the idea that first of all an optimal
mean field, which assures a stable ground state for the QRPA vibrations,
should be constructed by involving all types of
particle-particle like interaction.

\section{Conclusions}
In the previous sections we developed a new formalism for the
double beta decay. The QRPASMF formalism formulated here is
different from the standard $pnQRPA$ approach in that the static
ground state includes the correlations coming from the ph and pp
two-body interaction. In the previous work we showed that this
defines a stable ground state with respect to the RPA excitation.
Here we show that the new formalism provides a double beta
transition amplitude which does not vanish in a large interval
of the pp interaction strength. Moreover this amplitude is
diminished with respect to that predicted by the $pnQRPA$ in the
stability interval of $\chi_1$, by a factor of 6.
Such a suppression is due to the fact that the second leg of the
double beta transition is very much quenched otherwise the
first leg being only slightly modified. As a matter of
fact this is an important result since it improves the agreement
with the experimental data for the half life by a factor of 36.
Therefore the present formalism produces a moderate suppression
of the double beta transition amplitude. Moreover the predicted
amplitude is not very insensitive to increasing the strength
$\chi_1$ of the pp interaction.  
This seems at variance with previous theoretical studies which 
considered the particle-particle interaction
very important for describing quantitatively the decay rate
in $pnQRPA$ approach, because it was not included in the self-consistent 
mean field.  The predicted result for the transition amplitude agrees 
quite well to the exact result. The exact transition amplitude
exhibits a minimum, but does not vanish, for the value of the
parameter  $\chi_1$ where the ground state energy of the odd-odd system 
gets lower than the ground state of the
mother nucleus. It is an open question whether this feature persists also
in realistic calculations and contributes to the vanishing of
the transition amplitude in the $pnQRPA$ formalism.

The Ikeda sum rule deviates from the N-Z value by an amount less
than $4\%$ for $\chi_1=0$ and $25\%$ at $\chi_1=1.$
A possible explanation for this discrepancy is that in the
single j level, the QRPA ground state of the mother nucleus
is $\beta^-$ decaying to only one state in the odd-odd nucleus.
 This feature contrasts to the exact calculations where there are three 
channels for the beta minus and two channels for the beta plus decay.

Concluding the results of the present paper constitute a very important
test of our idea that the suppression of the transition
amplitude with increasing strength $\chi_1$ of the particle-particle  
interaction is caused by a instability of the ground state. This un-wanted
feature can be cured by considering the contribution of the
particle-hole and particle-particle two-body interaction already
in the single particle mean field. 
The agreement of the present approach with the exact results
are encouraging us to apply  the proposed formalism to a
realistic two-body interaction and an extended model space for
the single particle motion.  
\section{Appendix A}
Here we give the analytical expressions for the quasiparticle
energies appearing in Eq. (3.2).
\bea
E_{11}&=&(\epsilon_p-\lambda_p)(|T_{p1}|^2-|T_{\widetilde{p1}}|^2)+
       (\epsilon_n-\lambda_n)(|T_{n1}|^2-|T_{\widetilde{n1}}|^2)
+\Delta_pT_{p1}T_{\widetilde{p1}}+\Delta_nT_{n1}T_{\widetilde{n1}}
\\ \nonumber
&&+\frac{2\beta}{\sqrt{2\Omega}}(T_{p1}T_{n1}-T_{\widetilde{p1}}
T_{\widetilde{n1}})
+\frac{2\Delta_{pn}}{\sqrt{2\Omega}}(T_{p1}T_{\widetilde{n1}}
+T_{\widetilde{p1}}T_{n1}),\\ \nonumber
E_{22}&=&(\epsilon_p-\lambda_p)(|T_{p2}|^2-|T_{\widetilde{p2}}|^2)+
       (\epsilon_n-\lambda_n)(|T_{n2}|^2-|T_{\widetilde{n2}}|^2)
+\Delta_pT_{p2}T_{\widetilde{p2}}+\Delta_nT_{n2}T_{\widetilde{n2}}
\\ \nonumber
&&+\frac{2\beta}{\sqrt{2\Omega}}(T_{p2}T_{n2}-T_{\widetilde{p2}}
T_{\widetilde{n2}})
+\frac{\Delta_{pn}}{\sqrt{2\Omega}}(T_{p2}T_{\widetilde{n2}}
+T_{\widetilde{p2}}T_{n2}),\\ \nonumber
E_{12}&=&(\epsilon_p-\lambda_p)(T_{p1}T_{p2}-T_{\widetilde{p1}} 
T_{\widetilde{p2}} )+
(\epsilon_n-\lambda_n)(T_{n1}T_{n2}-T_{\widetilde{n1}}T_{\widetilde{n2}}) 
\\ \nonumber
&&+\frac{1}{2}\Delta_p(T_{p1}T_{\widetilde{p2}}+ T_{p2}T_{\widetilde{p1}})+ 
\frac{1}{2}\Delta_n(T_{n1}T_{\widetilde{n2}}+T_{n2}T_{\widetilde{n1}})
\\ \nonumber
&&+\frac{\beta}{\sqrt{2\Omega}}(T_{p1}T_{n2}+T_{p2}T_{n1}-T_{\widetilde{p1}}
T_{\widetilde{n2}}-T_{\widetilde{p2}}T_{\widetilde{n1}}  )
\\ \nonumber
&&+\frac{\Delta_{pn}}{\sqrt{2\Omega}}(T_{p1}T_{\widetilde{n2}}+
T_{p2}T_{\widetilde{n1}}+T_{\widetilde{p1}}T_{n2}
+T_{\widetilde{p2}}T_{n1} ),\\ \nonumber
E_{21}&=&E_{12}.
\eea

\section{Appendix B}
By elementary calculations one finds the explicit expression for
the matrix elements $A_{ik}$ and $B_{ik}$ involved in the
linearized equations of motion (3.6). They are listed below:

\bea
A_{11}&=&2E_{11}+4\chi(T_{\widetilde{p1}}^2T_{n1}^2+T_{p1}^2T_{\widetilde{n1}}^2)
-4\chi_1(T_{\widetilde{n1}}^2T_{\widetilde{p1}}^2+T_{p1}^2T_{n1}^2),
\nonumber  \\
A_{12}&=&\sqrt{2}E_{21}+2\sqrt{2}\chi
(T_{n1}^2T_{\widetilde{p1}}T_{\widetilde{p2}}+
T_{p1}^2T_{\widetilde{n1}}T_{\widetilde{n2}}
+T_{\widetilde{p1}}^2T_{n1}T_{n2}+T_{\widetilde{n1}}^2T_{p1}T_{p2})
\nonumber\\
&&-2\sqrt{2}\chi_1(T_{\widetilde{n1}}^2T_{\widetilde{p1}}T_{\widetilde{p2}}
+T_{\widetilde{p1}}^2T_{\widetilde{n1}}T_{\widetilde{n2}}
+T_{p1}^2T_{n1}T_{n2}+T_{n1}^2T_{p1}T_{p2}), \nonumber\\
A_{13}&=&4\chi(T_{\widetilde{p1}}T_{\widetilde{p2}}T_{n1}T_{n2}+
T_{p1}T_{p2}T_{\widetilde{n1}}T_{\widetilde{n2}})
\nonumber \\
&-&4\chi_1 (T_{p1}T_{n1}T_{p2}T_{n2}+T_{\widetilde{p1}}T_{\widetilde{n1}}
T_{\widetilde{p2}}T_{\widetilde{n2}}),\nonumber\\
A_{21}&=&A_{12},\nonumber\\
A_{22}&=&E_{11}+E_{22}+2\chi
[(T_{\widetilde{p2}}T_{n1}+T_{\widetilde{p1}}T_{n2})^2
+(T_{\widetilde{n2}}T_{p1}+T_{\widetilde{n1}}T_{p2})^2]
\nonumber\\
&&-2\chi_1[(T_{\widetilde{p2}}T_{\widetilde{n1}}+
T_{\widetilde{p1}}T_{\widetilde{n2}})^2
+(T_{n2}T_{p1}+T_{n1}T_{p2})^2], \nonumber\\
A_{23}&=&\sqrt{2}E_{21}+2\sqrt{2}\chi(T_{\widetilde{p2}}^2T_{n1}T_{n2}
+T_{\widetilde{n2}}^2T_{p1}T_{p2}+T_{n2}^2T_{\widetilde{p1}}T_{\widetilde{p2}}
+T_{p2}^2T_{\widetilde{n1}}T_{\widetilde{n2}}) \nonumber\\
&&-2\sqrt{2}\chi_1(T_{\widetilde{n2}}^2T_{\widetilde{p1}}T_{\widetilde{p2}}
+T_{\widetilde{p2}}^2T_{\widetilde{n1}}T_{\widetilde{n2}}+
T_{n2}^2T_{p1}T_{p2}+
T_{p2}^2T_{n1}T_{n2}),  \nonumber\\
A_{31}&=&A_{13}, \nonumber\\
A_{32}&=&A_{23},\nonumber\\
A_{33}&=&2E_{22}+4\chi(T_{\widetilde{p2}}^2T_{n2}^2+T_{p2}^2T_{\widetilde{n2}}^2)
-4\chi_1(T_{\widetilde{p2}}^2T_{\widetilde{n2}}^2+T_{p2}^2T_{n2}^2).
\eea
\bea
B_{11}&=&8(\chi+\chi_1)T_{\widetilde{p1}}T_{\widetilde{n1}}T_{p1}T_{n1},
\nonumber\\
B_{12}&=&2\sqrt{2}(\chi+\chi_1)(T_{\widetilde{p1}}T_{\widetilde{n2}}T_{p1}T_{n1}+
T_{\widetilde{p1}}T_{\widetilde{n1}}T_{p2}T_{n1}+
T_{\widetilde{n1}}T_{\widetilde{p2}}T_{p1}T_{n1}+
T_{\widetilde{n1}}T_{\widetilde{p1}}T_{p1}T_{n2}), \nonumber\\
B_{13}&=&4\chi(T_{\widetilde{p1}}T_{\widetilde{n2}}T_{p2}T_{n1}+
T_{\widetilde{n1}}T_{\widetilde{p2}}T_{p1}T_{n2})
+4\chi_1(T_{\widetilde{p1}}T_{\widetilde{n1}}T_{n2}T_{p2}+
T_{\widetilde{p2}}T_{\widetilde{n2}}T_{n1}T_{p1}),\nonumber\\
B_{21}&=&B_{12}\nonumber\\
B_{22}&=&2\chi\left((T_{\widetilde{p2}}T_{n1}+T_{\widetilde{p1}}T_{n2})
(T_{\widetilde{n2}}T_{p1}+T_{\widetilde{n1}}T_{p2})
+(T_{\widetilde{n2}}T_{p1}+T_{\widetilde{n1}}T_{p2})
(T_{\widetilde{p2}}T_{n1}+T_{\widetilde{p1}}T_{n2})\right )
\nonumber\\
&&+2\chi_1\left((T_{\widetilde{p2}}T_{\widetilde{n1}}+
T_{\widetilde{p1}}T_{\widetilde{n2}})
(T_{n2}T_{p1}+T_{n1}T_{p2})+
(T_{n2}T_{p1}+T_{n1}T_{p2})
(T_{\widetilde{p2}}T_{\widetilde{n1}}+
T_{\widetilde{p1}}T_{\widetilde{n2}})\right ),\nonumber\\
B_{23}&=&2\sqrt{2}(\chi+\chi_1)\nonumber \\
&&\left[
(T_{\widetilde{p2}}T_{\widetilde{n2}} (T_{n1}T_{p2}
+T_{p1}T_{n2})+T_{p2}T_{n2}
(T_{\widetilde{p1}}T_{\widetilde{n2}}+ T_{\widetilde{p2}}T_{\widetilde{n1}})
\right], \nonumber \\
B_{31}&=&B_{13},\nonumber\\
B_{32}&=&B_{23}, \nonumber\\
B_{33}&=&8(\chi+\chi_1)T_{\widetilde{p2}}T_{\widetilde{n2}}T_{p2}T_{n2}.
\eea
\section{Appendix C}
To calculate the matrix elements of H in the basis (4.1) we need
to know the result of acting with H on a given representative
state. To give an example let us consider the proton-proton
pairing interaction and determine the vector expansion:
\be
-\frac{G}{4}A^{\dagger}_{pp}A_{pp}|l\rangle_k=
\sum_{m}{\cal P}_{lm}|m\rangle_k,~k=gm,m,i,d.
\ee
Once the expansion coefficients are known, the matrix elements of
the pairing interaction are readily obtained:
\be
-\frac{G}{4}\langle
n|A^{\dagger}_{pp}A_{pp}|l\rangle_k=\sum_{m}O^{(k)}_{nm}{\cal P}_{lm}. 
\ee
The expansion coefficients associated to various two-body terms
of the model Hamiltonian considered for a single j-shell are
listed in tables 4-7, for the grand mother, mother, intermediate
and daughter systems respectively.
Using the expansion coefficients and the overlap matrix we
derived analytical expressions for the matrices ${\bar H}$
corresponding to the basis labeled by $gm$, $m$, $i$, $d$,
respectively. Further, these matrices are treated as described in
Section IV. 

\newpage

\begin{table}[h]
\caption{The basis states describing the nuclei "grand mother"
$|gm\rangle$, mother $|m\rangle$, intermediate $|i\rangle$ and
daughter $|d\rangle$ in a
single j shell.  }
\label{table. 1}
\begin{tabular}{lcccc}
&$\hskip0.5cm$(N,Z)=(13,3)$\hskip0.5cm$&$\hskip0.5cm$
(N,Z)=(12,4)
$\hskip0.5cm$&$\hskip0.5cm$(N,Z)=(11,5)$\hskip0.5cm$ 
&$\hskip0.5cm$(N,Z)=(10,6)$\hskip0.5cm$ \\
&  $x$=gm  &  $x$=m  &  $x$=i  &  $x$=d \\ \hline
$|1\rangle_x$ &$ |0,5,3\rangle $ &$ |2,6,0\rangle $ &$ |2,5,1\rangle $ 
&$ |3,5,0\rangle $\\
$|2\rangle_x$ &$ |1,6,1\rangle$ &$ |1,5,2\rangle$ &$ |1,4,3\rangle $ 
&$ |2,4,2\rangle$\\
$|3\rangle_x$ &               & $|0,4,4\rangle$ &$ |0,3,5\rangle $ 
&$ |1,3,4\rangle 
$\\ 
$|4\rangle_x$ & & & & $|0,2,6\rangle $ \\ 
\end{tabular}
\end{table}

\begin{table}[h]
\caption{The values of $({\cal N}^{(x)}_k)^{-1}/a$ with
$a=(30\times{10!})^{\frac{1}{2}}$. The norm of the state $|k\rangle _x$ is
denoted by ${\cal N}^{(x)}_k$. The index "x" labels the $0^+$ states in grand
mother (gm), mother (m), intermediate (i) and daughter (d)
nuclei, respectively}
\label{table 2} 
\begin{tabular}{lcccc}
$x$&$\hskip1cm$$|1\rangle_x$$\hskip1cm$&$\hskip1cm$$|2\rangle_x$$\hskip1cm$
&$\hskip1cm$$|3\rangle_x$$\hskip1cm$&$\hskip1cm$$|4\rangle_x$$\hskip1cm$\\
gm&$2^7\times3$ & $2^8\times 3 \times\sqrt{2}$&   &     \\
m &$2^9\times 3 \times\sqrt{5}$&$2^7\times\sqrt{13}$ & $2^6\times
3 \times \sqrt{\frac{11}{5}}$ &  \\   
i & $2^9\times \sqrt{3}$ & $2^7\times \sqrt{\frac{23}{5}}$ & $
2^4\times \sqrt{286}$ &  \\ 
d & $2^{10}\times 3$ & $\frac{2^8}{5}\times\sqrt{58}$ & $
\frac{2^5}{5}\times \sqrt{1122}$& $2^4\times \sqrt{429}$ \\ 
\end{tabular}
\end{table}

\begin{table}[h]
\caption{The overlaps of states characterizing the grand mother
(gm), mother (m), intermediate odd-odd (i) and daughter (d)
nuclei.}
\label{table.3}[h] 
\begin{tabular}{llcccc}
Nucleus& &$\hskip0.8cm$$|1\rangle$$\hskip0.8cm$ & $\hskip0.8cm$
$|2\rangle$$\hskip0.8cm$&$\hskip0.8cm$$|3\rangle$$\hskip0.8cm$&
$\hskip0.8cm$$|4\rangle$$\hskip0.8cm$\\ 
 & $|1\rangle$& 1 & -$\frac{\sqrt{2}}{3}$& &  \\
gm & $|2\rangle$& -$\frac{\sqrt{2}}{3}$& 1& &   \\ \hline  
&$|1\rangle$& 1 &$ -\frac{3}{\sqrt{65}}$&
$\frac{1}{3\sqrt{11}}$& \\ 
m&$|2\rangle$&$ -\frac{3}{\sqrt{65}}$& 1 &
$-3\sqrt{\frac{5}{143}}$ &   \\
&$|3\rangle$&$\frac{1}{3\sqrt{11}}$&$-3\sqrt{\frac{5}{143}}$ & 1
&   \\  \hline  
&$|1\rangle$& 1 &$ -2\sqrt{\frac{5}{69}}$&
$\frac{5}{\sqrt{858}}$&  \\
i&$|2\rangle$&$ -2\sqrt{\frac{5}{69}}$& 1 &
$-\sqrt{\frac{110}{299}}$ &   \\
&$|3\rangle$&$\frac{5}{\sqrt{858}}$&$-\sqrt{\frac{110}{299}}$ & 1
& \\ \hline 
 &$|1\rangle$& 1 &$ -\frac{3}{\sqrt{58}}$&
$\frac{2}{3}\sqrt{\frac{6}{187}}$&$-\frac{1}{6}\sqrt{\frac{3}{143}}$ \\ \\
&$|2\rangle$&$ -\frac{3}{\sqrt{58}}$& 1 &
$-\frac{53}{22}\sqrt{\frac{33}{493}}$ &$\frac{1}{2}\sqrt{\frac{66}{377}}$  \\ \\
d&$|3\rangle$&$\frac{3}{3}\sqrt{\frac{6}{187}}$&
$-\frac{53}{22}\sqrt{\frac{33}{493}}$ & 1
&$-\sqrt{\frac{13}{34}}$  \\
&$|4\rangle$     &$-\frac{1}{6}\sqrt{\frac{3}{143}}$
&$\frac{1}{2}\sqrt{\frac{66}{377}}$&
$-\sqrt{\frac{13}{34}}$& 1 \\  
\end{tabular}
\end{table}

\begin{table}[t]
\caption{The actions of various terms of the model Hamiltonian
on the states describing the grand mother nucleus (N,Z)=(13,3).
} 
\label{table.4}
\begin{tabular}{lcc}
 $O|n\rangle_{gm}$  & $\hskip1.5cm$ $|1\rangle_{gm}$ $\hskip1.5cm $& $\hskip1.5cm$ 
$|2\rangle_{gm}$ $\hskip1.5cm$ \\   \hline
$ A^\dagger_{pp}A_{pp}|1\rangle_{gm}$&     &
$-6\frac{{\cal N}^{(gm)}_1}{{\cal N}^{(gm)}_2}$    \\  
$A^\dagger_{pp}A_{pp}|2\rangle_{gm}$& & 36    \\ 
$A^\dagger_{nn}A_{nn}|1\rangle_{gm}$& 60&$-6\frac{{\cal
N}^{(gm)}_1}{{\cal N}^{(gm)}_2} $   \\ 
$A^\dagger_{nn}A_{nn}|2\rangle_m$& & 96        \\ 
$B^\dagger_{pn}B_{pn}|1\rangle_{gm}$& 36 &$6\frac{{\cal
 N}^{(gm)}_1}{{\cal N}^{(gm)}_2}$  \\ 
$ B^\dagger_{pn}B_{pn}|2\rangle_{gm}$& $
24\frac{{\cal N}^{(gm)}_2}{{\cal N}^{(gm)}_1}$
 & 18  \\ 
$A^\dagger_{pn}A_{pn}|1\rangle_{gm}$&24 &  \\ 
$A^\dagger_{pn}A_{pn}|2\rangle_{gm}$&$-24\frac{{\cal
 N}^{(gm)}_2}{{\cal N}^{(gm)}_1}$& 6 \\
\end{tabular}
\end{table}

\newpage
\begin{table}[t]
\caption{The actions of various terms of the model Hamiltonian
on the states describing the mother nucleus (N,Z)=(12,4).
}
\label{table.5}
\begin{tabular}{lccc}
$O|n\rangle_m$   &$\hskip1cm$  $|1\rangle_m$ $\hskip1cm$ 
&$\hskip1cm$ $|2\rangle_m$ $\hskip1cm$
& $\hskip1cm$ $|3\rangle_m$ $\hskip1cm$  \\ \hline
 $A^{\dagger}_{pp}A_{pp}|1\rangle_m$& 72     &     &     \\  
$A^{\dagger}_{pp}A_{pp}|2\rangle_m$& 
$-2\frac{{\cal N}^{(m)}_2}{{\cal N}^{(m)}_1}$& 32&       \\
$ A^\dagger_{pp}A_{pp}|3\rangle_m$& &$-12 \frac{{\cal
N}^{(m)}_3} {{\cal N}^{(m)}_2}$   &    \\ 
$ A^\dagger_{nn}A_{nn}|1\rangle_m$& 120     &     &     \\ 
$A^\dagger_{nn}A_{nn}|2\rangle_m$&$ -2\frac{{\cal N}^{(m)}_2}{{\cal N}^{(m)}_1}$
& 60    &    \\  
$A^\dagger_{nn}A_{nn}|3\rangle_m$& &$ -12\frac{{\cal
N}^{(m)}_3}{{\cal N}^{(m)}_2}$ & 36   \\ 
$B^\dagger_{pn}B_{pn}|1\rangle_m$& 4 &$48\frac{{\cal
N}^{(m)}_1}{{\cal N}^{(m)}_2}$&      \\ 
$ B^\dagger_{pn}B_{pn}|2\rangle_m$& $2\frac{{\cal N}^{(m)}_2}{{\cal N}^{(m)}_1}$
 & 30    &$20\frac{{\cal N}^{(m)}_2}{{\cal N}^{(m)}_3} $    \\
$B^\dagger_{pn}B_{pn}|3\rangle_m$& & $12\frac{{\cal
N}^{(m)}_3}{{\cal N}^{(m)}_2}$ & 40   \\ 
$A^\dagger_{pn}A_{pn}|1\rangle_m$& &$ -48\frac{{\cal
N}^{(m)}_1}{{\cal N}^{(m)}_2}$&      \\ 
$ A^\dagger_{pn}A_{pn}|2\rangle_m$&      & 14    &
$-20\frac{{\cal N}^{(m)}_2}{{\cal N}^{(m)}_3}   $  \\
 $A^\dagger_{pn}A_{pn}|3\rangle_m$&      &     &  36 \\
\end{tabular}
\end{table}

\begin{table}[t]
\caption{The actions of various terms of the model Hamiltonian
on the states describing the intermediate odd-odd nucleus (N,Z)=(11,5).
} 
\label{table.6}
\begin{tabular}{lccc}
$O|n\rangle_i$ & $\hskip1cm$ $|1\rangle_i$ $\hskip1cm$ & $\hskip1cm$ $|2\rangle_i$ 
$\hskip1cm$ &$\hskip1cm$ $|3\rangle_i$ $\hskip1cm$\\  \hline
$ A^\dagger_{pp}A_{pp}|1\rangle_i$& 64     &     &     \\  
$A^\dagger_{pp}A_{pp}|2\rangle_i$&$-6\frac{{\cal N}^{(i)}_2}
{{\cal N}^{(i)}_1}$  & 28& \\    
$A^\dagger_{pp}A_{pp}|3\rangle_i$& & 
$-20 \frac{{\cal N}^{(i)}_3}{{\cal N}^{(i)}_2}$ &    \\ 
$ A^\dagger_{nn}A_{nn}|1\rangle_i$& 100     &     &     \\ 
$A^\dagger_{nn}A_{nn}|2\rangle_i$&$ -6\frac{{\cal N}^{(i)}_2}{{\cal N}^{(i)}_1}$
& 64    &    \\ 
$A^\dagger_{nn}A_{nn}|3\rangle_i$& &$ -20\frac{{\cal
N}^{(i)}_3}{{\cal N}^{(i)}_2}$ & 36   \\ 
$B^\dagger_{pn}B_{pn}|1\rangle_i$& 20&$ 40\frac{{\cal
N}^{(i)}_1}{{\cal N}^{(i)}_2}$&      \\ 
$ B^\dagger_{pn}B_{pn}|2\rangle_i$&$ 6\frac{{\cal N}^{(i)}_2}{{\cal N}^{(i)}_1}$
 & 38    &$16\frac{{\cal N}^{(i)}_2}{{\cal N}^{(i)}_3} $    \\
$B^\dagger_{pn}B_{pn}|3\rangle_i$& &$ 20\frac{{\cal
N}^{(i)}_3}{{\cal N}^{(i)}_2}$ & 40   \\ 
$A^\dagger_{pn}A_{pn}|1\rangle_i$&6 &$ -40\frac{{\cal
N}^{(i)}_1}{{\cal N}^{(i)}_2}$&      \\ 
$ A^\dagger_{pn}A_{pn}|2\rangle_i$& &24    &
$-16\frac{{\cal N}^{(i)}_2}{{\cal N}^{(i)}_3}  $   \\
$ A^\dagger_{pn}A_{pn}|3\rangle_i$&      &     & 50   \\
\end{tabular}
\end{table}

\begin{table}[t]
\caption{The actions of various terms of the model Hamiltonian
on the states describing the daughter nucleus (N,Z)=(10,6)
} 
\label{table.7}
\begin{tabular}{lcccc}
 $O|n\rangle_d$  & $\hskip1cm$ $|1\rangle_d$ $\hskip1cm$ & 
$\hskip1cm$ $|2\rangle_d$ $\hskip1cm$
 &$\hskip1cm$  $|3\rangle_d$ $\hskip1cm$ &$\hskip1cm$ $|4\rangle_d$ 
$\hskip1cm$\\  \hline
$ A^\dagger_{pp}A_{pp}|1\rangle_d$& 96    &     &   &    \\ 
$A^\dagger_{pp}A_{pp}|2\rangle_d$& 
$-2\frac{{\cal N}^{(d)}_2}{{\cal N}^{(d)}_1}$& 56&    &      \\
$A^\dagger_{pp}A_{pp}|3\rangle_d$& & $-12 
\frac{{\cal N}^{(d)}_3}{{\cal N}^{(d)}_2}$ & 24 &   \\ 
 $A^\dagger_{pp}A_{pp}|4\rangle_d$&    &     &$ -30
\frac{{\cal N}^{(d)}_4}{{\cal N}^{(d)}_3}$  &   \\ 
$ A^\dagger_{nn}A_{nn}|1\rangle_d$& 120     &     &    &    \\ 
$A^\dagger_{nn}A_{nn}|2\rangle_d$&$ -2\frac{{\cal N}^{(d)}_2}{{\cal N}^{(d)}_1}$
& 80    &    &    \\ 
$A^\dagger_{nn}A_{nn}|3\rangle_d$& &$ -12\frac{{\cal
N}^{(d)}_3}{{\cal N}^{(d)}_2}$ & 48  &     \\
$A^\dagger_{nn}A_{nn}|4\rangle_d$& & &$-30\frac{{\cal
N}^{(d)}_4}{{\cal N}^{(d)}_3}$ & 24  \\ 
$B^\dagger_{pn}B_{pn}|1\rangle_d$& 6 &$60\frac{{\cal
 N}^{(d)}_1}{{\cal N}^{(d)}_2}$&     &     \\  
 $B^\dagger_{pn}B_{pn}|2\rangle_d$&$ 2\frac{{\cal N}^{(d)}_2}{{\cal N}^{(d)}_1}$
 & 32    &$32\frac{{\cal N}^{(d)}_2}{{\cal N}^{(d)}_3}$    &     \\
$B^\dagger_{pn}B_{pn}|3\rangle_d$& &$ 12\frac{{\cal
N}^{(d)}_3}{{\cal N}^{(d)}_2}$ & 42   & $ 12\frac{{\cal
N}^{(d)}_3}{{\cal N}^{(d)}_4}  $ \\ 
$B^\dagger_{pn}B_{pn}|4\rangle_d$& &  &  $30\frac{{\cal
 N}^{(d)}_4}{{\cal N}^{(d)}_3}$ & 36 \\ 
$A^\dagger_{pn}A_{pn}|1\rangle_d$& &$ -60\frac{{\cal
 N}^{(d)}_1}{{\cal N}^{(d)}_2}$&    &    \\  
$ A^\dagger_{pn}A_{pn}|2\rangle_d$&      &  14 &$-32\frac{{\cal
 N}^{(d)}_2}{{\cal N}^{(d)}_3}$ & \\ 
$ A^\dagger_{pn}A_{pn}|3\rangle_d$&      & &$ 36$  &$ -12\frac{{\cal
 N}^{(d)}_1}{{\cal N}^{(d)}_2}$ \\ 
$ A^\dagger_{pn}A_{pn}|4\rangle_d$& & &     
 & 66   \\  
\end{tabular}
\end{table}

\begin{figure}[h]
\centerline{\psfig{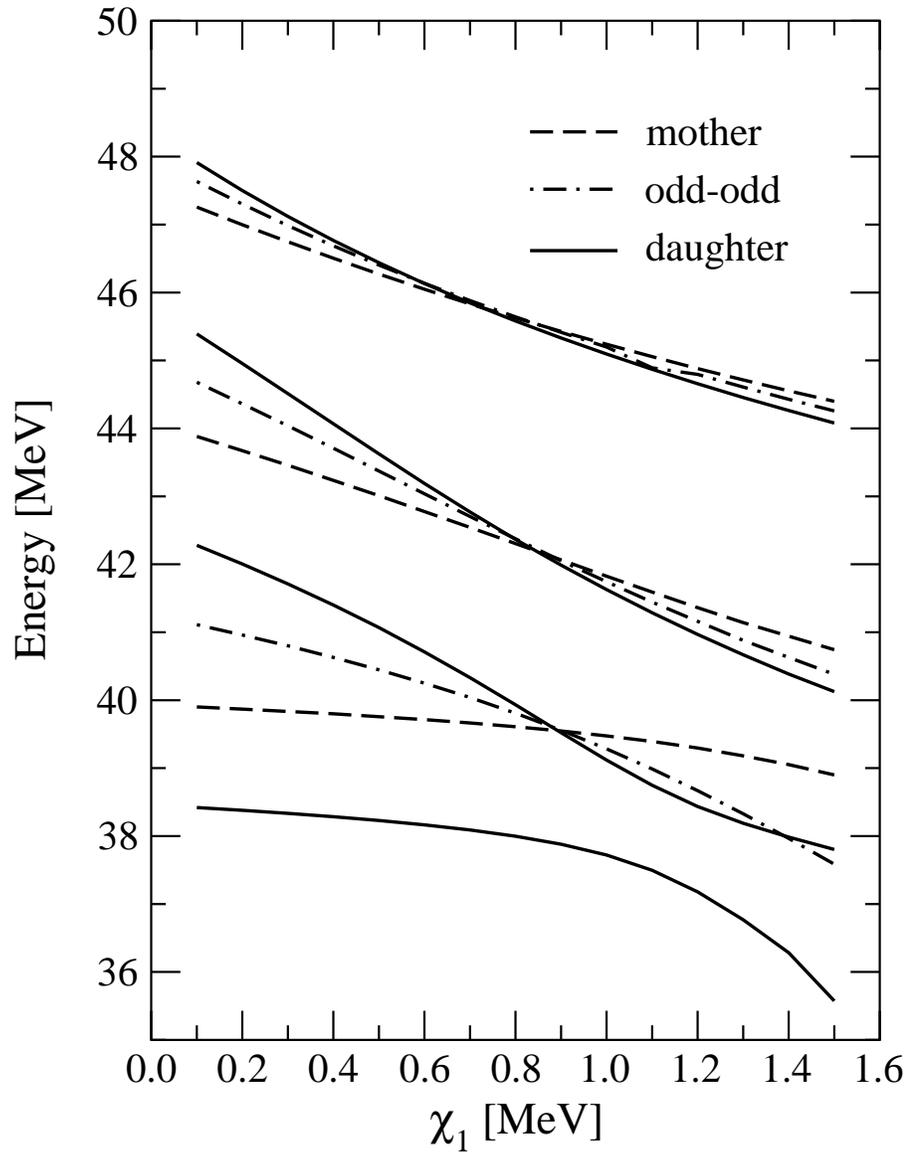}}
\caption{The exact energies for the states of the mother (dashed
line) intermediate (dashed dot line) and daughter (full line)
are plotted as function of the particle-particle interaction
strength, $\chi_1$.}
\label{Fig. 1}
\end{figure}
\begin{figure}[h]
\centerline{\psfig{figure=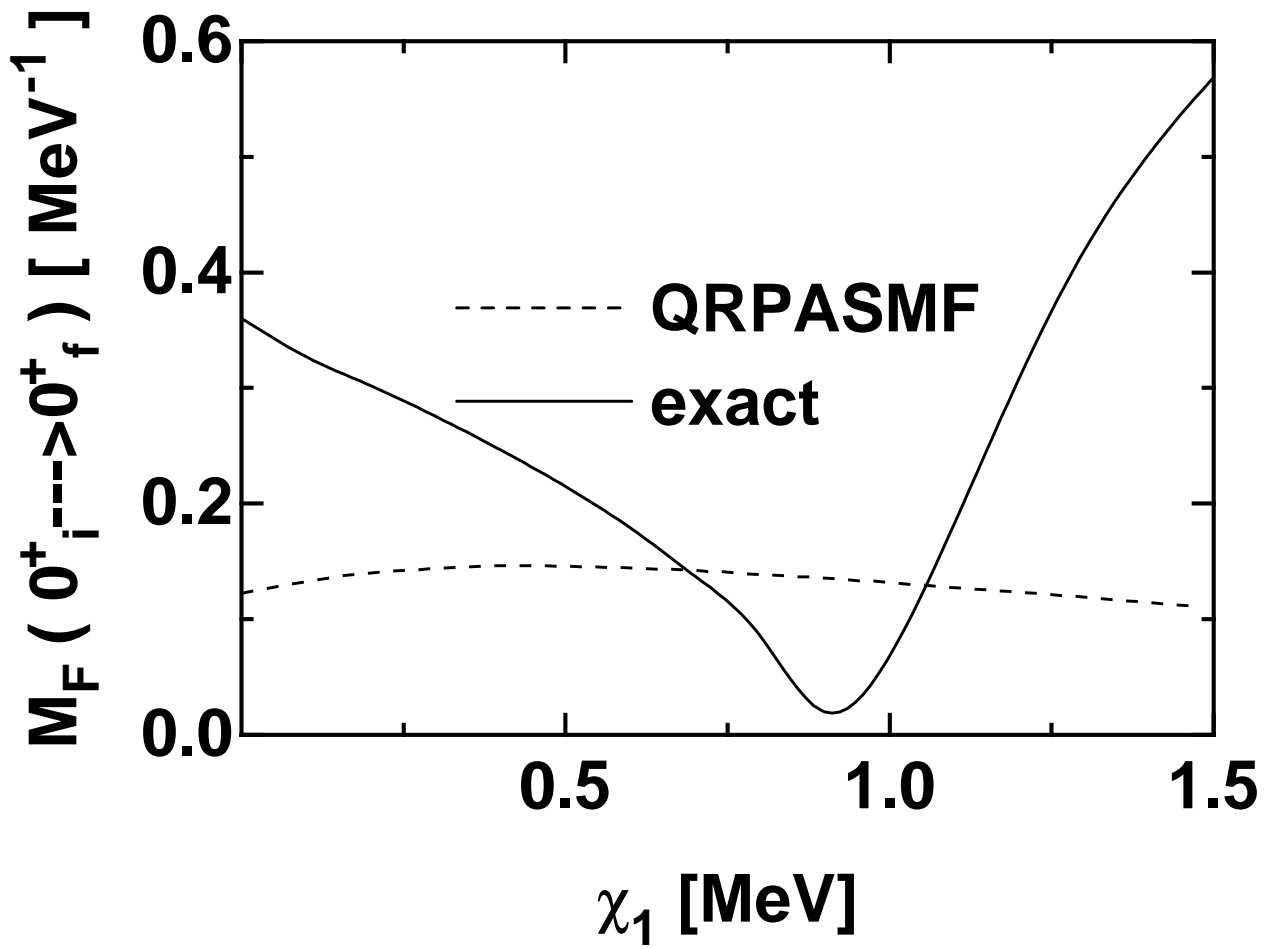,width=12cm,bbllx=5cm,%
bblly=7cm,bburx=18cm,bbury=26cm,angle=0}}
\caption{The transition amplitudes for the double beta
($2\nu\beta\beta$) Fermi transition, corresponding to the exact
eigenstates and energies of the model Hamiltonian (full line) and the new
QRPA approach of the present paper(dashed line), are plotted as
function of $\chi_1$. }
\label{Fig. 2}
\end{figure}
\begin{figure}[h]
\centerline{\psfig{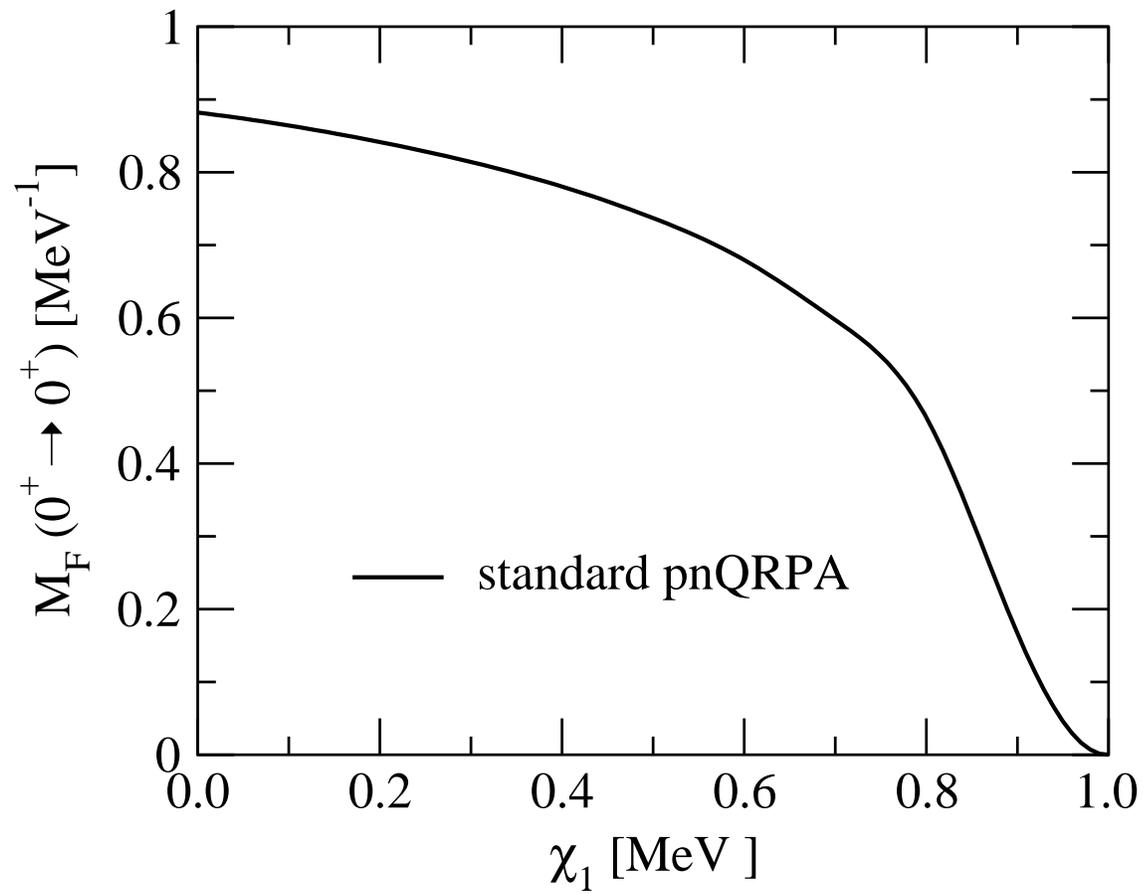}}
\caption{The transition amplitude for the double beta Fermi
transition calculated within the standard $pnQRPA$, is represented
as function of $\chi_1$.}
\label{Fig. 3}
\end{figure}
\begin{figure}[h]
\centerline{\psfig{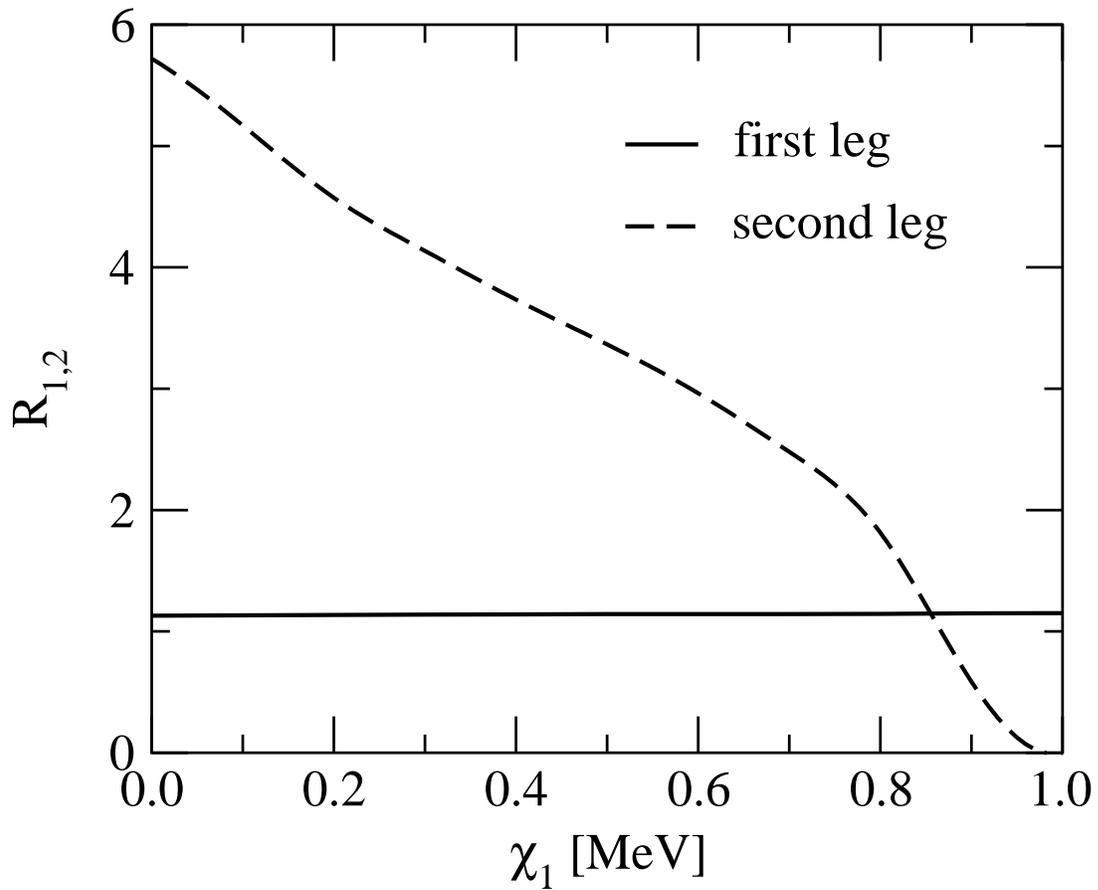}}
\caption{The ratio of the first single beta matrix element
involved in eq (5.2) obtained within the standard pnQRPA and the
present approach respectively (full line), is represented
as function of $\chi_1$. The ratio corresponding to the second
leg of the double beta transition is plotted by a dashed line. }
\label{Fig. 4}
\end{figure}
\begin{figure}[h]
\centerline{\psfig{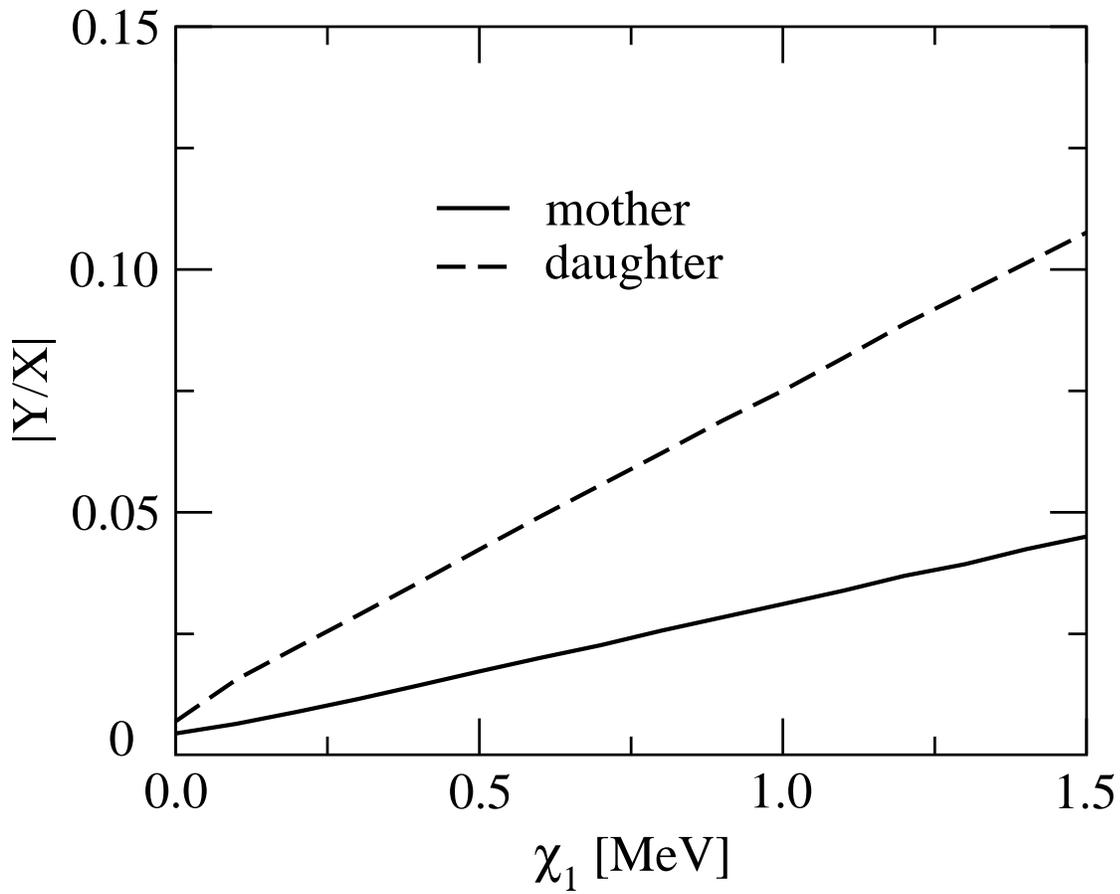}}
\caption{The magnitude of the ratio of the phonon amplitudes calculated within
the present approach, is represented
as function of $\chi_1$ for mother (full line) and daughter
nuclei (dashed line).}
\label{Fig. 5}
\end{figure}
\begin{figure}[h]
\centerline{\psfig{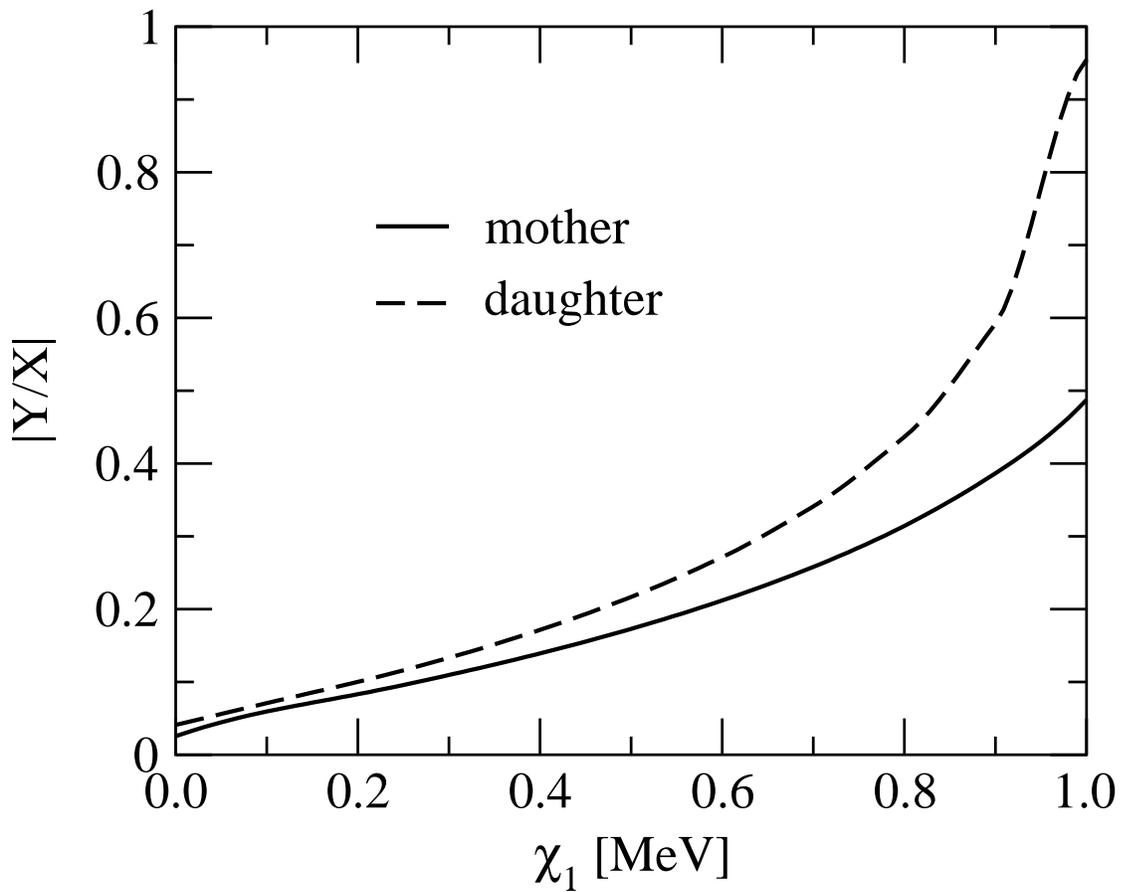}}
\caption{The magnitude of the ratio of the phonon amplitudes, calculated within
the standard $pnQRPA$, is represented
as function of $\chi_1$ for mother (full line) and daughter
nuclei (dashed line).}
\label{Fig. 6}
\end{figure}
\begin{figure}[h]
\centerline{\psfig{figure=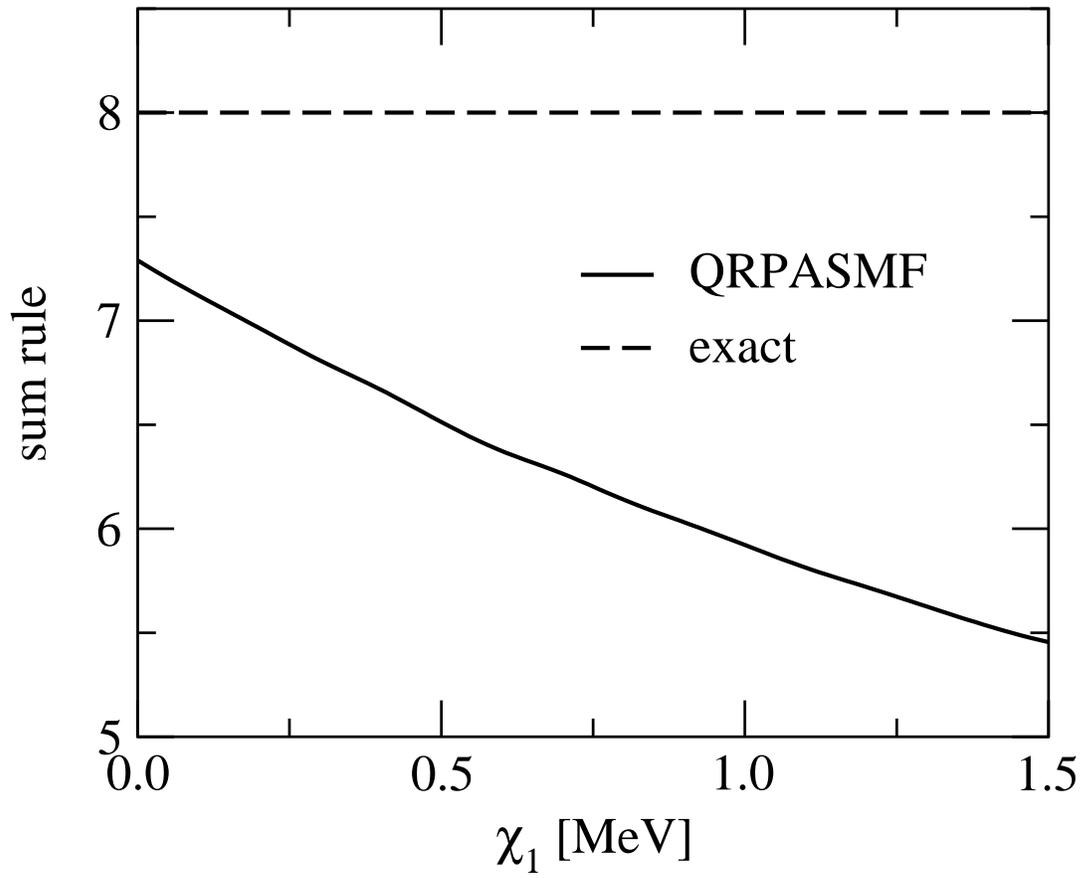,width=15cm,bbllx=5cm,%
bblly=10cm,bburx=18cm,bbury=26cm,angle=0}}
\caption{The $N-Z$ sum rule corresponding to the exact
description of nuclear states(dashed line) an to the present
approach, is represented
as function of $\chi_1$.}
\label{Fig. 7}
\end{figure}
\end{document}